\documentclass[sigconf,10pt]{acmart}
\usepackage{listings}
\usepackage{booktabs} 
\usepackage{algorithm2e}
\usepackage{natbib}
\usepackage{multirow}

\setcopyright{none}
\usepackage{url}

\begin{document}

\title{Umbrella: A Unified Software Defined Development Framework}

\author{Douglas Comer}
\affiliation{%
  \institution{Purdue University, Computer Science Department}
  \city{West Lafayette} 
  \state{IN} 
  \postcode{47907}
}
\email{comer@cs.purdue.edu}

\author{Rajas H. Karandikar}
\affiliation{%
  \institution{Purdue University, Computer Science Department}
  \city{West Lafayette} 
  \state{IN} 
  \postcode{47907}
}
\email{rkarandi@purdue.edu }

\author{Adib Rastegarnia}
\affiliation{%
  \institution{Purdue University, Computer Science Department}
  \city{West Lafayette} 
  \state{IN} 
  \postcode{47907}
}
\email{arastega@purdue.edu}

\renewcommand\footnotetextcopyrightpermission[1]{} 
\pagestyle{plain} 


\begin{abstract}
The \emph{Northbound (NB)} APIs that SDN controllers provide differ in terms of architecture, syntax, naming convention, data resources, and usage.  Using NB APIs to write SDN applications makes each application dependent on the API of a specific controller. To bring NB APIs from different vendors under one umbrella and make programming of SDN applications independent of specific controllers, we propose a unified software defined development framework that we call \emph{Umbrella}. This paper presents the key components of the software and reports some preliminary results. 
\end{abstract}

%



\maketitle

\section{Introduction}
Software defined networking (SDN) is an emerging trend for future design of Internet management systems that breaks vertical integration by decoupling the control plane from data plane and providing flexibility that allows software to program the data plane hardware directly\cite{Nunes2014}. In the current SDN paradigm, SDN controllers compromise three key layers including the data plane, control plane, and application layers. Most of the SDN controllers employ two \emph{Application Programming Interfaces (APIs)}, known the \emph{Northbound (NB)} and \emph{Southbound} APIs. Network applications use NB APIs to communicate with the controller, specify network behavior, define configuration requirements, and program forwarding devices.  The NB APIs offered by SDN controllers such as ONOS \cite{onos}, OpenDayLight \cite{odl} differ.  Even two REST APIs may differ in terms of syntax, naming convention, data resources, and usage. Currently, an SDN application depends on the NB API an SDN controller offers. Even for simple SDN applications, some of the modules used to collect topology information, generate and install flow rules, monitor topology changes, and collect flow rule statistics must be recoded from scratch when switching an application from one SDN controller to another. The Open Networking Foundation (ONF) started an NB API working group to provide a set of standard NB APIs at multiple levels of abstraction. Unfortunately, the effort has not produced widely-accepted standardized NB APIs.

We take a new approach to creating a standardized programming interface by creating new management abstractions and then providing a way to map the abstractions onto heterogeneous NB APIs.  We call our unified development framework \emph{Umbrella}. We use an architecture that follows the approach used in operating systems. An operating system provides a high-level I/O abstraction for applications, and uses a set of device drivers to map the abstractions into hardware commands suitable for a device from a given vendor. Our architecture takes the same approach by dividing the development framework into two conceptual parts: a module that provides a high-level, controller-independent NB API abstractions, and a set of controller-specific
translation modules that map the abstractions into NB API requests and commands that are suitable for various SDN controllers. Our design goals are: 

\begin{itemize}
\item Design and implement a development framework that provides a new set of abstractions for SDN applications, keeping the abstractions independent of the NB APIs used by specific SDN controllers.
\item Design and implement a set of modules that use the proposed abstractions to provide information needed by SDN applications, such as topology, network statistics, and real time topology changes. 
\item Increase portability of SDN applications across SDN controllers, and make it easy for a programmer to evaluate a specific application on multiple SDN controllers (e.g., to compare performance).
\item Provide a software defined network programming framework that reduces programming complexity, allows a programmer to write SDN applications without requiring a programmer to master low-level details for each SDN controller, and avoids locking an application to a specific controller.
\end{itemize}

\section{An Overview of Umbrella Architecture}
\begin{figure}[ht]
\centering
\includegraphics[scale=0.5]{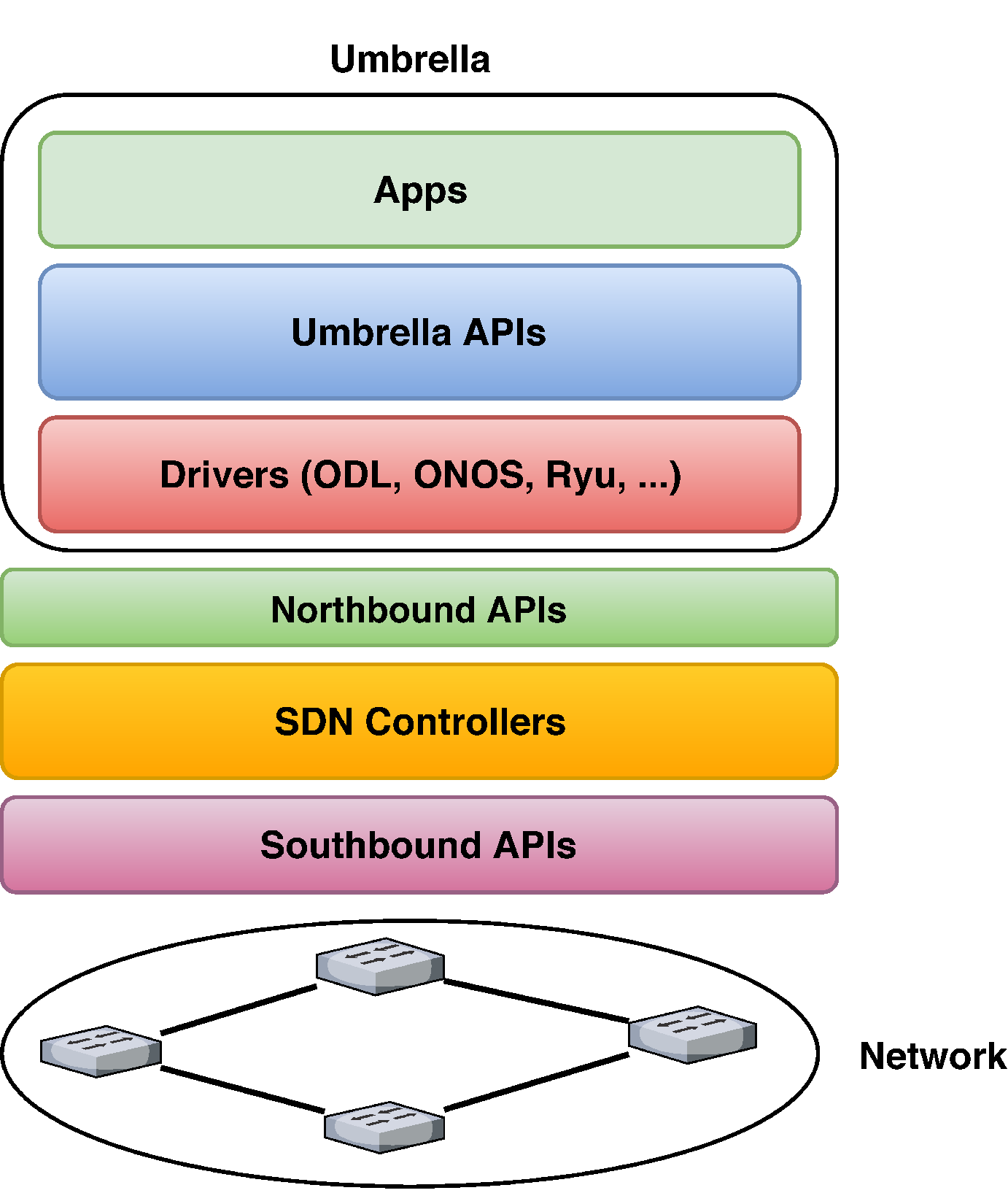}
\caption{Umbrella Architecture}
\label{umbrella}
\end{figure}

Figure \ref{umbrella} illustrates the architecture of Umbrella. 

Umbrella consists of three key components: 
\begin{itemize}
\item \emph{Umbrella APIs}: Umbrella provides a set of high-level and generic APIs that programmers use to write SDN applications. The APIs include abstractions that allow applications to install, and remove flow rules, retrieve topology information, monitor topology changes, retrieve network statistics and a list of installed flow rules, compute end-to-end paths between network end points, and employ custom path finding algorithms.

\item \emph{Drivers}: A set of drivers translate between Umbrella's high-level APIs and the NB API of specific controllers; when an application expresses a request or command, a driver translates into the controller-specific equivalent, and sends the result to the controller to be executed. 

\item \emph{Apps}: Programmers use Umbrella APIs to write portable, controller-independent SDN applications.  If new controllers appear or if the NB API used by a controller changes, a programmer can write a new driver module for the controller or modify an existing driver.
\end{itemize}

\section{Experimental Results}
\begin{figure}[!ht]
\centering
\includegraphics[scale=0.6]{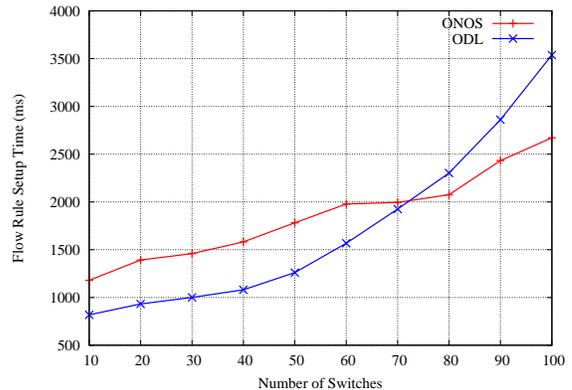}
\caption{Flow Rule Setup Time}
\label{fig2}
\end{figure}

We wrote driver modules for the ONOS and ODL SDN controllers.  We then used the Umbrella high-level API to write a controller-independent application that installs one-directional flow rules to forward traffic between a sender and a receiver. We ran a script that performs a total of 10 experiments using Mininet with linear topology of size 10, 20, 30... 100. Each experiment is performed 5 times and includes the following steps: 1) Create a Mininet instance to setup the topology with a specific size. 2) Have the sender transmit packets at the rate of 1 pkt/ms, and after 2 seconds, have our application install flow rules. 3) Arrange for the receiver (i.e. the destination host in the topology) to receive packets throughout the experiment and to report the number of packets received.  Packet loss is then used to compute the flow rule setup time. We illustrate the average flow rule setup time vs number of switches in the topology for ODL and ONOS controllers in Figure \ref{fig2}. As the results show, the flow rule setup time increases as the number of switches increases between two end points. In addition, the simulation results show that ODL outperforms ONOS in smaller network topologies However, ONOS has a better performance in terms of flow rule installation time in larger network topologies when compared with ODL.  Our framework makes comparing controller performance straightforward.

\label{er}

\section{Conclusion and Future work}
In this work-in-progress paper, we present a unified software development framework that can be used to implement SDN applications independent of NB APIs that different SDN controllers provide.  We plan to complete the framework by adding more drivers to support more types of NB APIs from additional SDN controllers. We will also add more reusable modules, such as a monitoring module; additional modules will be helpful in designing and implementing of complex SDN applications.   
\label{conclusion}
\bibliographystyle{ACM-Reference-Format}
\bibliography{sample-bibliography} 

\end{document}